\documentclass[aps,showpacs,preprintnumbers,amsmath,amssymb,prc]{revtex4}

\usepackage{graphicx}
\usepackage{dcolumn}
\usepackage{bm}

\newcommand{\epsfigbox}[5]{%
\begin{figure} \vspace{#3}%
\includegraphics[width=10.0cm]{#2}%
\caption{ \label{fig:#1} #5} \vspace{#4}
\end{figure}}

\newcommand{\epsfigpox}[5]{%
\begin{figure} \vspace{#3}%
\includegraphics[width=11.0cm]{#2}%
\caption{ \label{fig:#1} #5} \vspace{#4}
\end{figure}}

\newcommand{\quarterthin}{\kern 0.0417em}

\newcommand{\bra}[1]{\langle#1}
\newcommand{\ket}[1]{#1\rangle}

\begin{document}

\title{Systematics of $g$ factors of $2_1^+$ states in even-even nuclei from Gd to Pt: \\
A microscopic description by the projected shell model}

\author{
Bao-An Bian$^{1,2}$, Yao-Min Di$^{1}$, Gui-Lu Long$^{3,1}$, Yang
Sun$^{4,1}$, Jing-ye Zhang$^5$, Javid A. Sheikh$^6$}

\affiliation{
$^{1}$Department of Physics, Xuzhou Normal
University, Xuzhou, Jiangsu 221009, P.R. China \\
$^{2}$Institute of Low Energy Nuclear Physics, Beijing Normal
University, Beijing 100875, P. R. China \\
$^{3}$Department of Physics, Tsinghua University, Beijing 100084,
P.R. China \\
$^{4}$Department of Physics and Joint Institute for Nuclear
Astrophysics, University of Notre Dame, Notre Dame, Indiana
46556, USA \\
$^{5}$Department of Physics and Astronomy, University of
Tennessee, Knoxville, Tennessee 37996, USA \\
$^{6}$Department of Physics, University of Kashmir, Hazrathbal,
Srinagar, Kashmir 190 006, India }

\begin{abstract}
The systematics of $g$ factor of first excited $2^+$ state vs
neutron number $N$ is studied by the projected shell model. The
study covers the even-even nuclei of all isotopic chains from Gd
to Pt. $g$ factors are calculated by using the many-body
wavefunctions that reproduces well the energy levels and $B(E2)$s
of the ground-state bands. For Gd to W isotopes the characteristic
feature of the $g$ factor data along an isotopic chain is
described by the present model. Deficiency of the model in the $g$
factor description for the heavier Os and Pt isotopes is
discussed.
\end{abstract}

\date{\today}
\maketitle

\section{Introduction}

Nuclear magnetic dipole moment can provide valuable information on
the microscopic structure of a nuclear system. It is a sensitive
probe of nuclear wavefunctions and hence can serve as a strict
testing ground for theoretical models. Because of the
intrinsically opposite signs of the neutron and proton $g_s$, a
study of the gyromagnetic factor ($g$ factor) enables
determination of the detailed structure for underlying states. For
example, variation of $g$ factors often is a clear indicator for a
single-particle component that strongly influences the total
wavefunction. With advances in modern experimental techniques and
sensitive detectors, progress in $g$ factor measurement has
continuously been made. In a recent paper \cite{Berant04}, Berant
{\it et al.}, by summarizing their new and the accumulated data
from Refs. \cite{Data1,Data2}, raised an interesting question on
the systematic behavior of the first excited $2^+$ $g$ factors
(denoted as $g(2^+_1)$ hereafter) in the rare earth nuclei and the
heavier mass region. The data for these even-even nuclei indicate
characteristic features of the systematics (see Fig. 4 below):
with increasing neutron number, $g(2^+_1)$ factors display a
decreasing trend in the Gd, Dy, and Er isotopes; stay nearly
constant within a range of the Yb and Hf chains; then change to an
increasing trend in the W and Os isotopes; but show a flat
behavior in the Pt chain.

Clearly, the overall behavior of these $g$ factors exhibits a
large deviation from the rotor value, $Z/A$, which has only a very
weak and smooth dependence on nucleon numbers \cite{Greiner66}. On
the other hand, the proton-neutron version of the Interacting
Boson Model \cite{Sam84} gives an overly strong particle number
dependence, and thus fails to reproduce the flat behavior of the
$g$ factors near the midshell \cite{Berant04}. These facts may
suggest that in realistic nuclear systems, $g$ factors reflect a
delicate interplay between collective and single-particle degrees
of freedom, which is dictated by the detailed shell structure.
Very recently, Zhang {\it et al.} have discussed the systematical
behavior of these $g(2^+_1)$ factors in terms of a simple
phenomenological model \cite{Zhang06}. However, it is desired that
the $g(2^+_1)$ systematics can be described by microscopic
theories. Spherical shell model calculation is applicable only to
those heavy nuclei near the shell closures (for a recent example,
see Ref. \cite{Brown05}). There have been microscopic models
employed in the $g$ factor calculation for heavy, deformed nuclei
\cite{Tanabe88,Ansari90,Cescato91,Saha93,Sun94,Vela99,Sun01,Sun02}.
However, except in Ref. \cite{Saha93}, those calculations focused
mainly on one or a few chosen examples in an isotopic chain. A
microscopic description of the experimental $g(2^+_1)$ data for
the large mass region as presented in Ref. \cite{Berant04} remains
as a challenge to microscopic theories.

In the present article, we carry out a systematical study for
$g(2^+_1)$ factors. As a theoretical tool, we employ the code
developed in Ref. \cite{Sun94}, which is based on the Projected
Shell Model (PSM) \cite{PSM}. In the PSM approach, one introduces
axially deformed basis and performs exactly angular-momentum
projection on the intrinsic states from deformed mean-field
calculations. For even-even nuclei, angular-momentum projection on
the lowest $K = 0$ state generates a rotational band, which is the
main component of the ground-state band (g-band) including the
$2^+_1$ state of our interest. The lowest $K=0$ state is the
quasiparticle vacuum obtained microscopically through the Nilsson
+ BCS calculations in a large single-particle space. We thus
expect this model to be an appropriate microscopic theory for the
present investigation. The calculation is performed for nuclei
from the Gd ($Z$ = 64) to the Pt ($Z$ = 78) isotopic chains, with
neutron numbers ranging from $N$ = 88 to 120. The nuclei studied
here are known to have very different collective properties; for
example, they can be well-deformed, less-deformed, or soft nuclei.
The $g$ factor of the first $2^+$ state is dependent on details of
the total wavefunction. The deviations from the collective $Z/A$
trend are mainly understood as a consequence of single-particle
make-up of the wavefunction and the interplay between the
collective degree of freedom and single particles. The present
model employs deformed Nilsson single-particle states at fixed
deformation. As we shall discuss, this scheme works well for
well-deformed nuclei but is insufficient for description of soft
nuclei in the heavier Os and Pt region.

\section{Outline of the theory}

A calculation for medium to heavy deformed nuclei in terms of the
conventional (spherical) shell model is not feasible despite
recent computational advances. The successful description of heavy
deformed nuclei can be traced back to the introduction of the
Nilsson potential \cite{Nilsson69}. In the Nilsson model, nuclear
states are described by considering nucleons moving in a deformed
potential. Deformed states are defined in the body-fixed frame of
reference in which the rotational symmetry is broken. In order to
calculate the observables, it is necessary to restore the broken
rotational symmetry in the wavefunction. This can be done by using
the standard angular momentum projection method.  The projected
states are then used to diagonalize a two-body shell model
Hamiltonian. Thus, our approach follows closely with the basic
philosophy of the conventional shell model. The main difference to
the conventional shell model is that in the PSM, one starts with a
deformed basis rather than a spherical one. For the details of the
projection method, we refer to the PSM review article \cite{PSM}
and references cited therein.

The PSM constructs the shell-model space by using the axially
symmetric Nilsson states with a quadrupole deformation
$\epsilon_2$. Pairing correlations are incorporated into the
Nilsson states by the BCS calculations. The consequence of the
Nilsson-BCS calculations defines a quasiparticle vacuum
$\left|\Phi(\epsilon_2)\right> \equiv \left|\Phi\right>$ and the
associated set of quasiparticle states in the intrinsic frame. For
the low-lying nuclear states near the ground state the PSM
wavefunction can be expressed as
\begin{equation}
\left| \Psi^I_M \right> = f^I_{K=0}\,\hat P^I_{MK=0}\left|\Phi
\right>,
\label{wf}
\end{equation}
which is the angular-momentum-projected quasiparticle vacuum state
with $f$ being the normalization factor. In Eq. (\ref{wf}), $\hat
P^I_{MK}$ is the angular-momentum-projection operator \cite{RS80}
\begin{equation}
\hat P^I_{MK} = {2I+1 \over 8\pi^2} \int d\Omega\,
D^{I}_{MK}(\Omega)\, \hat R(\Omega).
\end{equation}

As in the early PSM calculations, we use the pairing plus
quadrupole-quadrupole ($QQ$) Hamiltonian with inclusion of the
quadrupole-pairing term
\begin{equation}
\hat H = \hat H_0 - {1 \over 2} \chi \sum_\mu \hat Q^\dagger_\mu
\hat Q^{}_\mu - G_M \hat P^\dagger \hat P - G_Q \sum_\mu \hat
P^\dagger_\mu\hat P^{}_\mu . \label{hamham}
\end{equation}
In Eq. (\ref{hamham}), $\hat H_0$ is the spherical single-particle
Hamiltonian, which contains a proper spin--orbit force
\cite{Nilsson69}. As discussed in Ref. \cite{PSM}, the $QQ$-force
strength $\chi$ is adjusted such that the quadrupole deformation
$\epsilon_2$ is obtained as a result of the self-consistent
mean-field HFB calculation. The monopole pairing strength $G_M$ is
of the standard form $G_M = \left[20.12 \mp13.13(N-Z)/A\right]/A$,
with ``$-$" for neutrons and ``$+$" for protons, which
approximately reproduces the observed odd--even mass differences
in this mass region \cite{PSM}. The quadrupole pairing strength
$G_Q$ is assumed to be proportional to $G_M$, with the
proportionality constant being fixed to be 0.16 for all nuclei
considered in this paper.  The same constant has been used in the
previous calculations for rare earth nuclei \cite{PSM}. For the
valence single-particle space, we have included three major
shells, $N=4,5,6$ (3,4,5), for neutrons (protons).

In short, the procedure of the present calculation is that based
on a deformed Nilsson potential with pairing included in the BCS
treatment, one performs explicit angular-momentum projection with
a two-body interaction which conforms (through self-consistent
conditions) with the mean-field Nilsson potential. The Hamiltonian
with separable forces serves as an effective interaction, the
strengths of which have been fitted to experimental data.  The
deformed single-particle states with deformation parameters
$\epsilon_2$ are used solely as a starting basis.  It is
sufficient for a calculation to have these deformation parameters
close to the ``true" nuclear deformation. Of course, a large
departure from a true deformation would result in a significant
enhancement in dimension of the configuration space, with the
extreme case being the conventional shell model based on a
spherical basis $(\epsilon_2 \equiv 0)$.

\section{Energy levels and BE(2) values}

\begin{table}
\caption{The quadrupole deformation parameters with which the
deformed bases are constructed.} \label{tab:1}
\begin{tabular}{ccccccccccc}
\hline\hline\noalign{\smallskip}
Z = 64 & $^{152}$Gd & $^{154}$Gd & $^{156}$Gd & $^{158}$Gd & $^{160}$Gd \\
$\epsilon_2$ & 0.212 & 0.278 & 0.295 & 0.305 & 0.320 \\
\hline\noalign{\smallskip}
Z = 66 & $^{154}$Dy & $^{156}$Dy & $^{158}$Dy & $^{160}$Dy & $^{162}$Dy & $^{164}$Dy \\
$\epsilon_2$ & 0.200 & 0.240 & 0.260 & 0.270 & 0.280 & 0.290 \\
\hline\noalign{\smallskip}
Z = 68 & $^{156}$Er & $^{158}$Er & $^{160}$Er & $^{162}$Er & $^{164}$Er & $^{166}$Er & $^{168}$Er & $^{170}$Er \\
$\epsilon_2$ & 0.195 & 0.230 & 0.257 & 0.265 & 0.258 & 0.262 & 0.273 & 0.276 \\
\hline\noalign{\smallskip}
Z = 70 & $^{164}$Yb & $^{166}$Yb & $^{168}$Yb & $^{170}$Yb & $^{172}$Yb & $^{174}$Yb & $^{176}$Yb \\
$\epsilon_2$ & 0.245 & 0.250 & 0.260 & 0.265 & 0.269 & 0.266 & 0.263 \\
\hline\noalign{\smallskip}
Z = 72 & $^{166}$Hf & $^{168}$Hf & $^{170}$Hf & $^{172}$Hf & $^{174}$Hf & $^{176}$Hf & $^{178}$Hf & $^{180}$Hf \\
$\epsilon_2$ & 0.219 & 0.235 & 0.240 & 0.250 & 0.253 & 0.246 & 0.241 & 0.231 \\
\hline\noalign{\smallskip}
Z = 74 & $^{168}$W & $^{170}$W & $^{172}$W & $^{174}$W & $^{176}$W & $^{178}$W & $^{180}$W & $^{182}$W & $^{184}$W & $^{186}$W \\
$\epsilon_2$ & 0.193 & 0.201 & 0.217 & 0.220 & 0.225 & 0.195 & 0.190 & 0.195 & 0.195 & 0.190 \\
\hline\noalign{\smallskip}
Z = 76 & $^{178}$Os & $^{180}$Os & $^{182}$Os & $^{184}$Os & $^{186}$Os & $^{188}$Os & $^{190}$Os & $^{192}$Os \\
$\epsilon_2$ & 0.188 & 0.172 & 0.170 & 0.173 & 0.158 & 0.154 & 0.150 & 0.145 \\
\hline\noalign{\smallskip}
Z = 78 & $^{182}$Pt & $^{184}$Pt & $^{186}$Pt & $^{188}$Pt & $^{190}$Pt & $^{192}$Pt & $^{194}$Pt & $^{196}$Pt & $^{198}$Pt \\
$\epsilon_2$ & 0.197 & 0.187 & 0.175 & 0.135 & 0.128 & 0.116 & 0.113 & 0.120 & 0.170 \\
\hline\hline\noalign{\smallskip}
\end{tabular}
\end{table}

We study 61 nuclei from the Gd, Dy, Er, Yb, Hf, W, Os, and Pt
isotopic chains. This large group includes nuclei with very
different collective behavior. It is well-known that with neutron
number around 90, nuclei are traditionally known as $\gamma$-soft
nuclei. On the other hand, the heavier isotopes in the Os and Pt
chains contain also significant $\gamma$-softness, and these are
typical examples of $O(6)$ nuclei according to the Interacting
Boson Model \cite{IBM}. Between these two regions, nuclei are
strongly deformed, for most of which the deformation is axial and
the low-lying spectrum typically exhibits a rotor behavior. In
Table I, we list the deformation parameters with which the
deformed bases are constructed. The listed quadrupole deformations
$\epsilon_2$ agree with the systematic trend of those
experimentally adopted ones \cite{Raman} although the absolute
values of ours are smaller. Note that it is not necessary for our
input deformations to be exactly the same as those extracted from
experiment as long as the so-constructed bases can correctly
describe experimental $B(E2)$s (see Figs. 2 and 3 below).  In the
calculation, when the calculation condition is fixed we do not
have a freedom to readjust the parameters to reproduce the
g-factors. Namely, under a fixed calculation condition, $g$
factors are predicted and the underlying physics is discussed.

\epsfigbox{fg 1}{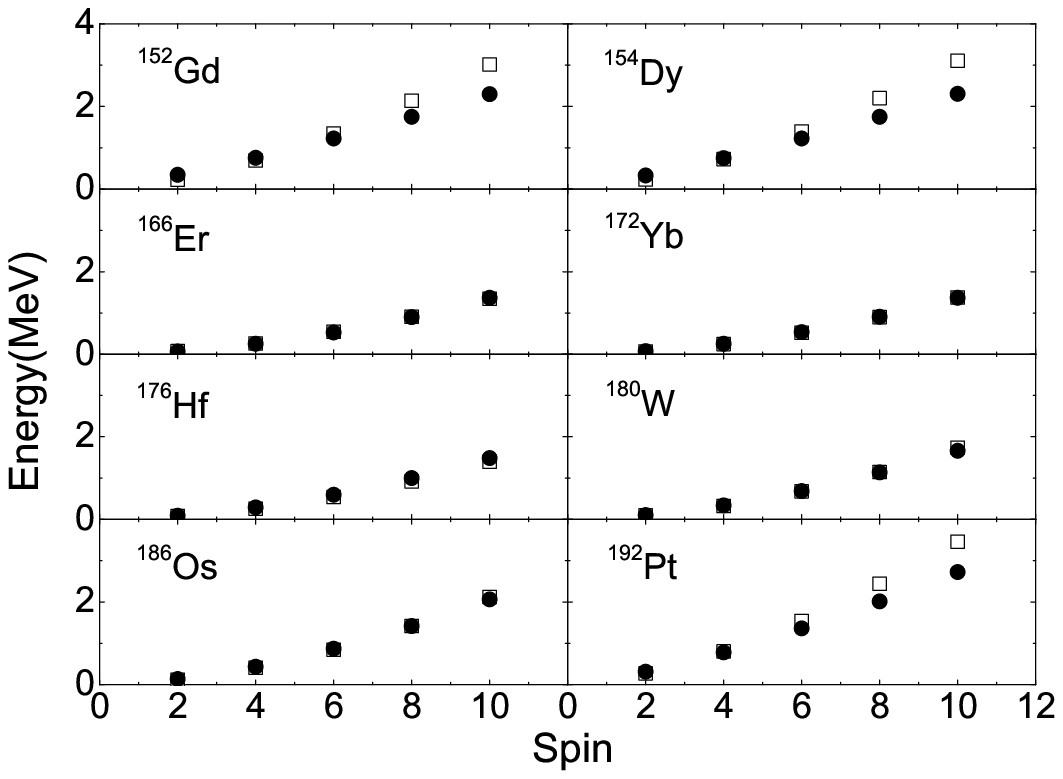}{0pt}{0pt} { Comparison of calculated
energies with experimental data for g-bands. Open squares
represent the calculated results and solid circles the data. }

Fig. 1 shows the calculated energy levels for g-bands together
with experimental data. We include in the figure only one nucleus
selected from each of the isotopic chains because all the
calculations have achieved the same level of agreement. These
examples are chosen to represent nuclei with distinct collective
behavior. For instance, with neutron number 88, $^{152}$Gd and
$^{154}$Dy are typical $\gamma$-soft nuclei lying in the
transitional region. $^{166}$Er, $^{172}$Yb, and $^{176}$Hf are
representative examples of strongly deformed nuclei lying in the
midshell, which have nearly constant moment of inertia. Finally,
$^{186}$Os and $^{192}$Pt are again $\gamma$-soft in the
transitional region. For all these nuclei with very different
rotational behavior, it can be seen from Fig. 1 that the energy
levels have been well reproduced by the calculation. Note that for
the strongly deformed nuclei lying in the midshell, the energy
intervals are considerably smaller, corresponding to larger
moments of inertia in these nuclei. The deviation seen in the
higher-spin states in $^{152}$Gd, $^{154}$Dy, and $^{192}$Pt is
caused by the g-band interactions with other configurations that
have not been considered in Eq. (\ref{wf}).

\epsfigbox{fg 2}{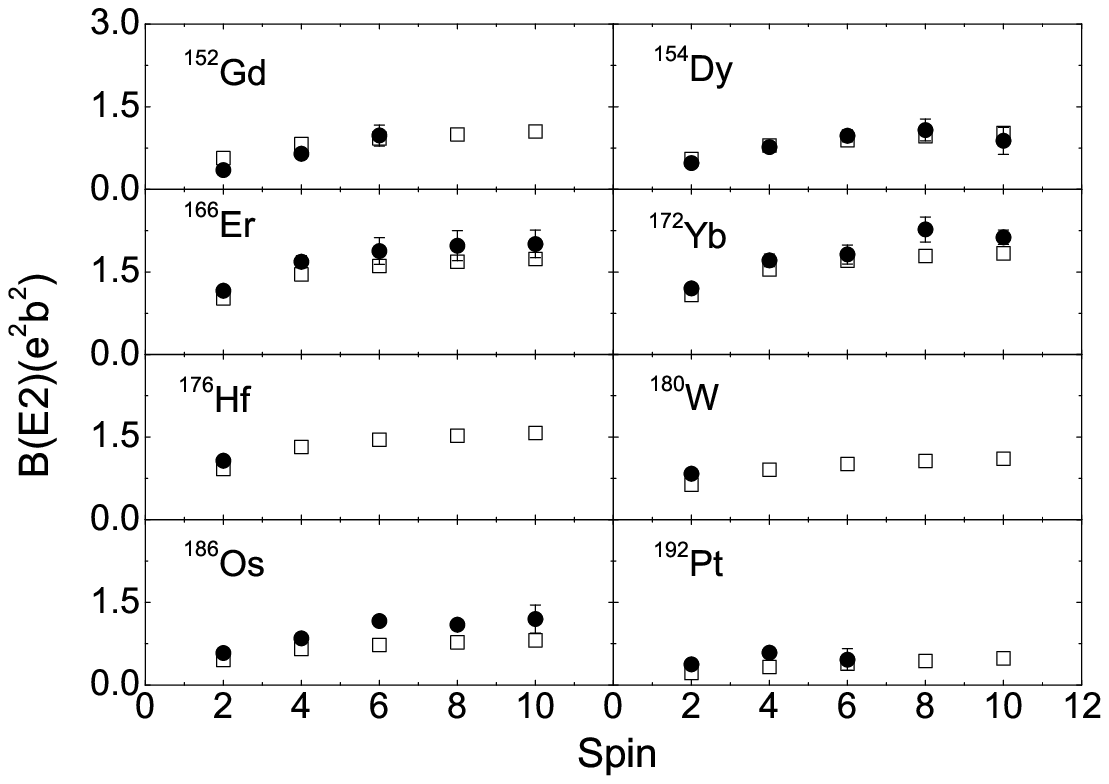}{0pt}{0pt} { Comparison of calculated
g-band $B(E2)$ values with available data. Open squares represent
the calculated results and solid circles (with error bars) the
data. }

We further calculate the g-band $B(E2)$ values for all nuclei
studied in this paper using the same deformation parameters listed
in Table I. The $B(E2)$ values that are related to an electric
transition probability from an initial state $I$ to a final state
$I-2$ are given by
\begin{equation}
B(E2, I\rightarrow I-2) = \frac {1}{2I + 1} \left| \right<
\Psi^{I-2} || \hat Q_2 ||\Psi^I \left> \right|^2,
\label{BE2}
\end{equation}
where wavefunctions $\left|\Psi^I\right>$ are those in Eq.
(\ref{wf}). The effective charges used in the calculation are the
standard ones $e^\pi=1.5e$ and $e^\nu=0.5e$. The effective charges
are fixed for all nuclei studied in this paper without any
individual adjustment. Any variations in the calculated $B(E2)$
values, among one rotational band or between those in different
nuclei, are subject to the structure change in wavefunctions. The
calculation is compared with available data in Fig. 2. Again, we
include in this figure only one nucleus selected from each of the
isotopic chains, the same set of nuclei as shown in Fig. 1. It can
be seen that the $B(E2)$ values are also nicely reproduced. Not
only are the absolute values in each nucleus correctly given, but
also the variations as a function of angular-momentum are
described. We note in particular that the $B(E2)$s in $^{192}$Pt
are well reproduced; as we shall see later, for the Pt isotopic
chain we encounter difficulties in the g-factor calculation.

\epsfigpox{fg 3}{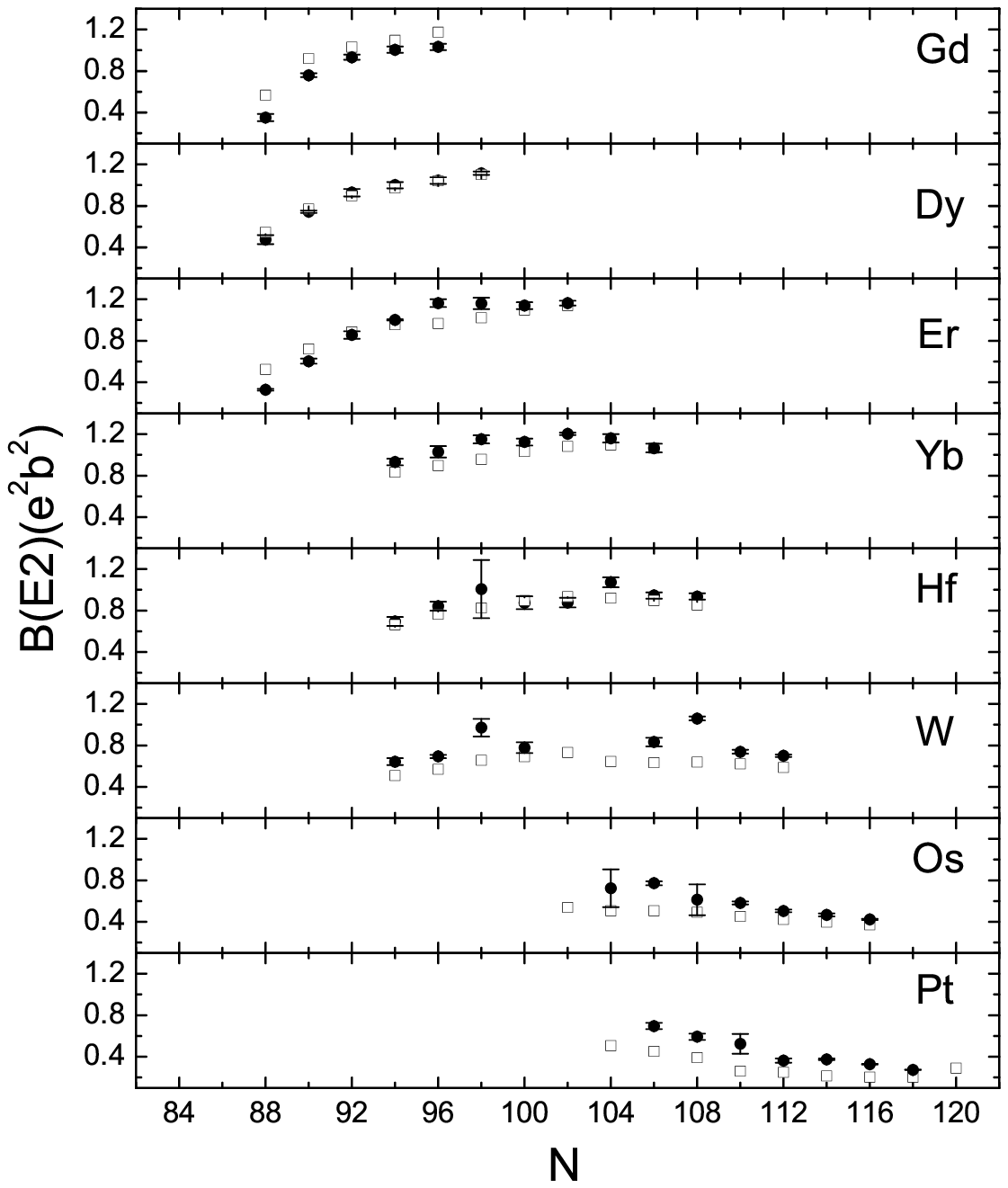}{0pt}{0pt} {Comparison of calculated
$B(E2$, $I=2\rightarrow I=0$) with available data. Open squares
represent the calculated results and solid circles (with error
bars) the data. }

\begin{table}
\caption{Comparison of calculated $B(E2$, $I=2\rightarrow I=0$)
(in $e^2b^2$) with available data.} \label{tab:2}
\begin{tabular}{ccccccccccc}
\hline\hline\noalign{\smallskip}
Z = 64 & $^{152}$Gd & $^{154}$Gd & $^{156}$Gd & $^{158}$Gd & $^{160}$Gd \\
Exp. & 0.35(3) & 0.76(2) & 0.93(3) & 1.01(3) & 1.03(3) \\
Th.  & 0.57    & 0.92    & 1.03    & 1.09    & 1.18    \\
\hline\noalign{\smallskip}
Z = 66 & $^{154}$Dy & $^{156}$Dy & $^{158}$Dy & $^{160}$Dy & $^{162}$Dy & $^{164}$Dy \\
Exp. & 0.48(4) & 0.75(1) & 0.93(4) & 1.00(3) & 1.04(3) & 1.11(2) \\
Th.  & 0.55    & 0.77    & 0.90    & 0.97    & 1.04    & 1.10    \\
\hline\noalign{\smallskip}
Z = 68 & $^{156}$Er & $^{158}$Er & $^{160}$Er & $^{162}$Er & $^{164}$Er & $^{166}$Er & $^{168}$Er & $^{170}$Er \\
Exp. & 0.33(1) & 0.60(3) & 0.86(4) & 1.00(5) & 1.16(8) & 1.16(5) & 1.14(3) & 1.16(2) \\
Th.  & 0.52    & 0.72    & 0.89    & 0.96    & 0.97    & 1.02    & 1.10    & 1.14    \\
\hline\noalign{\smallskip}
Z = 70 & $^{164}$Yb & $^{166}$Yb & $^{168}$Yb & $^{170}$Yb & $^{172}$Yb & $^{174}$Yb & $^{176}$Yb \\
Exp. & 0.93(3) & 1.03(5) & 1.15(4) & 1.12(3) & 1.20(1) & 1.16(4) & 1.07(4) \\
Th.  & 0.83    & 0.90    & 0.96    & 1.03    & 1.08    & 1.10    & 1.07    \\
\hline\noalign{\smallskip}
Z = 72 & $^{166}$Hf & $^{168}$Hf & $^{170}$Hf & $^{172}$Hf & $^{174}$Hf & $^{176}$Hf & $^{178}$Hf & $^{180}$Hf \\
Exp. & 0.69(4) & 0.84(4) & 1.01(30) & 0.88(6) & 0.88(5) & 1.07(5) & 0.95(3) & 0.94(3) \\
Th.  & 0.66    & 0.77    & 0.83     & 0.89    & 0.94    & 0.92    & 0.90    & 0.85    \\
\hline\noalign{\smallskip}
Z = 74 & $^{168}$W & $^{170}$W & $^{172}$W & $^{174}$W & $^{176}$W & $^{178}$W & $^{180}$W & $^{182}$W & $^{184}$W & $^{186}$W \\
Exp. & 0.64(3) & 0.69(2) & 0.97(9) & 0.78(5) &      &      & 0.83(4) & 1.06(2) & 0.74(2) & 0.70(1) \\
Th.  & 0.51    & 0.57    & 0.66    & 0.69    & 0.73 & 0.65 & 0.63    & 0.64    & 0.62    & 0.59    \\
\hline\noalign{\smallskip}
Z = 76 & $^{178}$Os & $^{180}$Os & $^{182}$Os & $^{184}$Os & $^{186}$Os & $^{188}$Os & $^{190}$Os & $^{192}$Os \\
Exp. &      & 0.72(18) & 0.77(2) & 0.61(15) & 0.58(2) & 0.51(1) & 0.47(1) & 0.42(1) \\
Th.  & 0.54 & 0.51     & 0.51    & 0.49     & 0.45    & 0.43    & 0.40    & 0.37    \\
\hline\noalign{\smallskip}
Z = 78 & $^{182}$Pt & $^{184}$Pt & $^{186}$Pt & $^{188}$Pt & $^{190}$Pt & $^{192}$Pt & $^{194}$Pt & $^{196}$Pt & $^{198}$Pt \\
Exp. &      & 0.70(3) & 0.59(3) & 0.53(9) & 0.36(2) & 0.38(1) & 0.33(1) & 0.27(1) &      \\
Th.  & 0.51 & 0.45    & 0.39    & 0.26    & 0.25    & 0.22    & 0.20    & 0.20    & 0.29 \\
\hline\hline\noalign{\smallskip}
\end{tabular}
\end{table}

In Fig. 3, we plot the calculated $B(E2$, $I=2\rightarrow I=0$)
values for all the 61 nuclei considered in this work, and compare
them with available data. The numbers used for the figure are
listed in Table II. We stress that the variations in $B(E2)$ along
each isotopic chain, i.e. a rapid increase up to $N\approx 94$,
the flat behavior for $96\le N \le 108$, and a decreasing trend
after the midshell, are correctly described. A few local
exceptions with rather large $B(E2)$ values in the data (such as
in $^{172}$W, $^{182}$W, and $^{182}$Os) can not be understood by
the present calculation. The global trend of the $B(E2)$ values
with exclusion of the Os and Pt chains has been discussed by a
simple one-parameter model \cite{Zhang06}.

The agreement of the calculated $B(E2$, $I=2\rightarrow I=0$)
values with data (Figs. 2 and 3) as well as the reproduction of
the g-band energies (Fig. 1) indicate that under the present
calculation conditions, we are able to describe the basic
structure quantities for these nuclei. As far as the low-lying
energy levels and $B(E2)$s are concerned, the method works well.
The systematical agreement between the calculated and experimental
$B(E2)$s (Fig. 3) justifies the use of the deformation parameters
in Table I.

\section{g-factor results and discussion}

We now turn our discussion to $g$ factors. In the PSM, $g$ factors
can be directly computed as
\begin{equation}
g(I) = \frac {\mu(I)}{\mu_N I} = \frac {1}{\mu_N I} \left[ \mu_\pi
(I) + \mu_\nu (I) \right],
\label{gfactor}
\end{equation}
with $\mu_\tau (I)$ being the magnetic moment of a state $\Psi^I$
\begin{eqnarray}
\mu_\tau (I) &=& \left< \Psi^I_{I} | \hat
\mu^\tau_z | \Psi^I_{I} \right> \nonumber\\
&=& {I\over{\sqrt{I(I+1)}}} \left<
\Psi^{I} || \hat \mu^\tau || \Psi^{I} \right> \nonumber\\
&=& \frac{I}{\sqrt{I(I+1)}} \left[
     g^{\tau}_l \langle \Psi^I || \hat j^\tau ||\Psi^I
     \rangle + (g^{\tau}_s - g^{\tau}_l)
     \langle \Psi^I || \hat s^\tau || \Psi^I \rangle \right]
     ,
\label{moment}
\end{eqnarray}
where $\tau = \pi$ and $\nu$ for protons and neutrons, respectively.

We use the same wavefunctions that are used to evaluate $B(E2)$
values. In the angular-momentum-projection theory, the reduced
matrix element for an operator $\hat m$ (with $\hat m$ being
either $\hat j$ or $\hat s$ in Eq. (\ref{moment})) can be
explicitly expressed as
\begin{eqnarray*}
\langle \Psi^{I} || \hat m^\tau || \Psi^{I} \rangle
 &=& \sum_{K_i,K_f} f_{K_i}^I f_{K_f}^I \sum_{M_i
, M_f , M} (-)^{I - M_f} \left(
\begin{array}{ccc}
I & 1 & I \\-M_f & M &M_i
\end{array} \right)
\bra \Phi | {\hat{P}^{I}}_{K_f M_f} \hat m_{1M} \hat{P}^{I}_{K_i
M_i} | \Phi \ket\nonumber \\
 &=& (2I+1) \sum_{K_i,K_f} (-)^{I-K_f} f_{K_i}^I f_{K_f}^I
 \sum_{M^\prime,M^{\prime\prime}} \left(
\begin{array}{ccc}
I & 1 & I \\-K_{f} & M^\prime & M^{\prime\prime}
\end{array} \right) \nonumber \\
 & & \times \int d\Omega {\it D}_{M'' K_{i}} (\Omega) \langle \Phi | \hat
m_{1M'} \hat{R}(\Omega) | \Phi \rangle .
\end{eqnarray*}
In our calculation, the following standard values for $g_l$ and
$g_s$ appearing in Eq. (\ref{moment}) are taken:
\begin{eqnarray*}
g_l^\pi &=& 1, ~~~~ g_s^\pi = 5.586 \times 0.75 ,\\
g_l^\nu &=& 0, ~~~~ g_s^\nu = -3.826 \times 0.75 .
\end{eqnarray*}
$g_s^\pi$ and $g_s^\nu$ are damped by a usual 0.75 factor from the
free-nucleon values to account for the core-polarization and
meson-exchange current corrections \cite{Castel90}. The same
values are used for all $g$ factor calculations in the present
paper, as in the previous projected shell model calculations,
without any adjustment for individual nuclei.

\epsfigpox{fg 4}{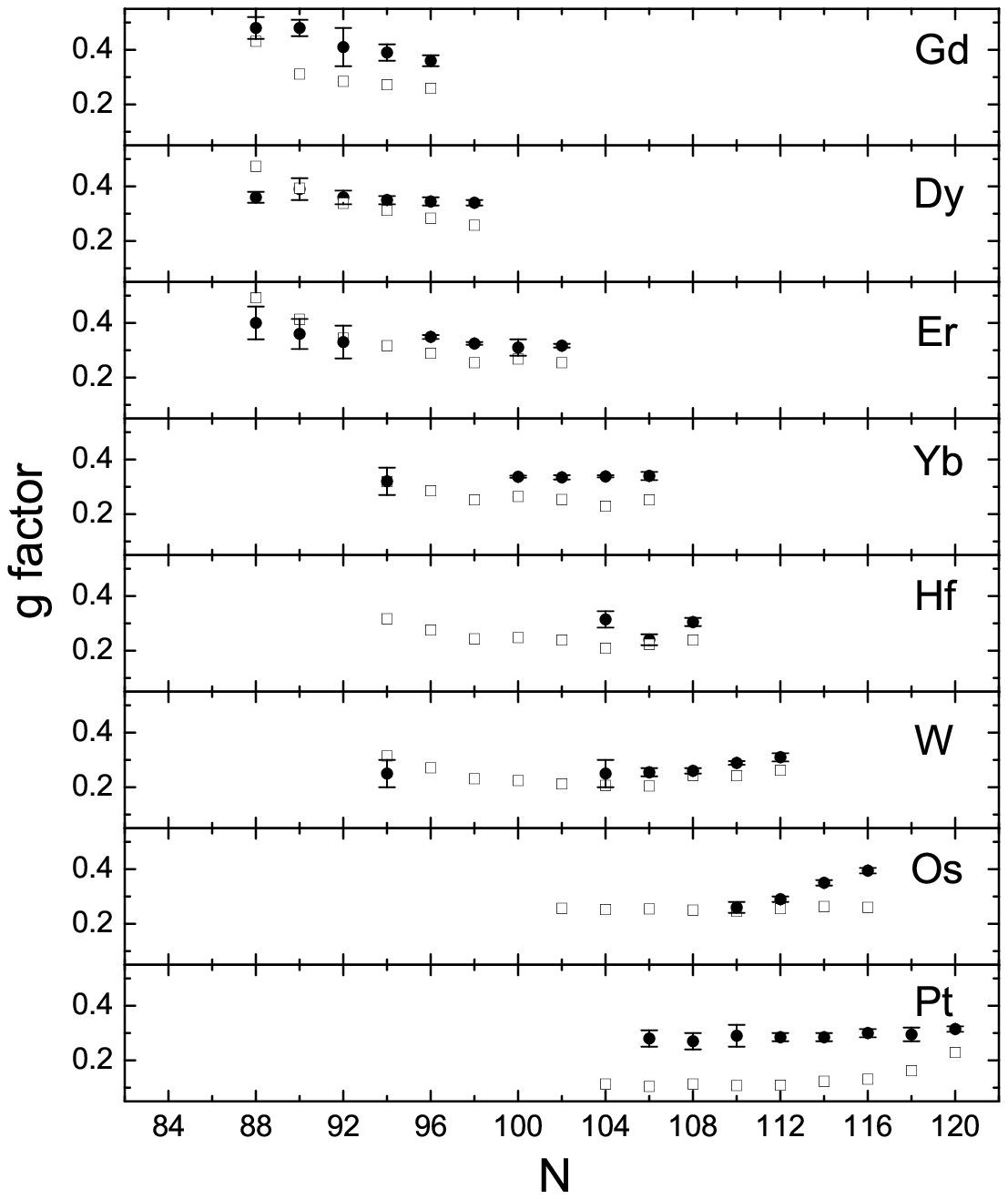}{0pt}{0pt} {Comparison of calculated
$2^+$ state $g$ factor with available data. Open squares represent
the calculated results and solid circles (with error bars) the
data. }

\begin{table}
\caption{Comparison of calculated $2^+$ state $g$ factor with
available data.} \label{tab:3}
\begin{tabular}{ccccccccccc}
\hline\hline\noalign{\smallskip}
$Z = 64$ & $^{152}$Gd & $^{154}$Gd & $^{156}$Gd & $^{158}$Gd & $^{160}$Gd \\
Exp. & 0.48(4) & 0.48(3) & 0.41(7) & 0.39(3) & 0.36(2) \\
Th.  & 0.43    & 0.31    & 0.29    & 0.27    & 0.26    \\
\hline\noalign{\smallskip}
$Z = 66$ & $^{154}$Dy & $^{156}$Dy & $^{158}$Dy & $^{160}$Dy & $^{162}$Dy & $^{164}$Dy \\
Exp. & 0.36(2) & 0.39(4) & 0.36(3) & 0.35(2) & 0.35(2) & 0.34(1) \\
Th.  & 0.47    & 0.39    & 0.34    & 0.31    & 0.28    & 0.26    \\
\hline\noalign{\smallskip}
$Z = 68$ & $^{156}$Er & $^{158}$Er & $^{160}$Er & $^{162}$Er & $^{164}$Er & $^{166}$Er & $^{168}$Er & $^{170}$Er \\
Exp. & 0.40(6) & 0.36(6) & 0.33(6) &      & 0.349(8) & 0.325(5) & 0.31(3) & 0.317(7) \\
Th.  & 0.49    & 0.41    & 0.35    & 0.32 & 0.29     & 0.25     & 0.27    & 0.26     \\
\hline\noalign{\smallskip}
$Z = 70$ & $^{164}$Yb & $^{166}$Yb & $^{168}$Yb & $^{170}$Yb & $^{172}$Yb & $^{174}$Yb & $^{176}$Yb \\
Exp. & 0.32(3) &      &      & 0.337(4) & 0.335(8) & 0.338(4) & 0.34(2) \\
Th.  & 0.32    & 0.29 & 0.25 & 0.26     & 0.25     & 0.23     & 0.25    \\
\hline\noalign{\smallskip}
$Z = 72$ & $^{166}$Hf & $^{168}$Hf & $^{170}$Hf & $^{172}$Hf & $^{174}$Hf & $^{176}$Hf & $^{178}$Hf & $^{180}$Hf \\
Exp. &      &      &      &      &      & 0.32(3) & 0.24(2) & 0.31(2) \\
Th.  & 0.32 & 0.28 & 0.24 & 0.25 & 0.24 & 0.21    & 0.22    & 0.24    \\
\hline\noalign{\smallskip}
$Z = 74$ & $^{168}$W & $^{170}$W & $^{172}$W & $^{174}$W & $^{176}$W & $^{178}$W & $^{180}$W & $^{182}$W & $^{184}$W & $^{186}$W \\
Exp. & 0.25(5) &      &      &      &      & 0.25(5) & 0.26(2) & 0.26(1) & 0.289(7) & 0.31(2) \\
Th.  & 0.32    & 0.27 & 0.23 & 0.22 & 0.21 & 0.21    & 0.20    & 0.24    & 0.24     & 0.26    \\
\hline\noalign{\smallskip}
$Z = 76$ & $^{178}$Os & $^{180}$Os & $^{182}$Os & $^{184}$Os & $^{186}$Os & $^{188}$Os & $^{190}$Os & $^{192}$Os \\
Exp. &      &      &      &      & 0.26(2) & 0.29(1) & 0.35(1) & 0.40(1) \\
Th.  & 0.26 & 0.26 & 0.25 & 0.25 & 0.25    & 0.26    & 0.26    & 0.26    \\
\hline\noalign{\smallskip}
$Z = 78$ & $^{182}$Pt & $^{184}$Pt & $^{186}$Pt & $^{188}$Pt & $^{190}$Pt & $^{192}$Pt & $^{194}$Pt & $^{196}$Pt & $^{198}$Pt \\
Exp. &      & 0.28(3) & 0.27(3) & 0.29(4) & 0.29(2) & 0.29(2) & 0.30(2) & 0.30(3) & 0.32(1) \\
Th.  & 0.11 & 0.11    & 0.11    & 0.11    & 0.11    & 0.12    & 0.13    & 0.16    & 0.23    \\
\hline\hline\noalign{\smallskip}
\end{tabular}
\end{table}

We present the systematics of $g$ factor of first 2$^+$ state for
all the isotopic chains from Gd to Pt. In Fig. 4, a comparison of
calculated results with available data \cite{Berant04,Data1,Data2}
is given. The numbers used for the figure are listed in Table III.
Interesting systematical features are clearly observed. The
experimental $g(2^+_1)$ values of Gd, Dy, and Er isotopes show a
downsloping trend with increasing neutron number $N$; those of
heavier W and Os isotopes exhibit an upsloping behavior; and,
according to the current set of data, the $g(2^+_1)$ factors of
Yb, Hf and Pt isotopes are almost constant within each isotopic
chain. From Fig. 4, it can be seen that the observed {\it trends}
for the isotopic chains from Gd to W are qualitatively reproduced
by the present calculation. Especially for most Dy, Er, and W
nuclei, a quantitative agreement with data is achieved. The
downsloping trend of Gd, Dy, and Er isotopes is well described.
The calculation further predicts that for the Yb, Hf, and W
isotopic chains, a downsloping trend is still visible for $N <
100$, but becomes nearly constant at the neutron midshell. The
upsloping trend of W isotopes with $N\ge 106$ is well reproduced
by the calculation. Nevertheless, there are cases where we find
disagreement. We shall comment on those cases later.

\epsfigbox{fg 5}{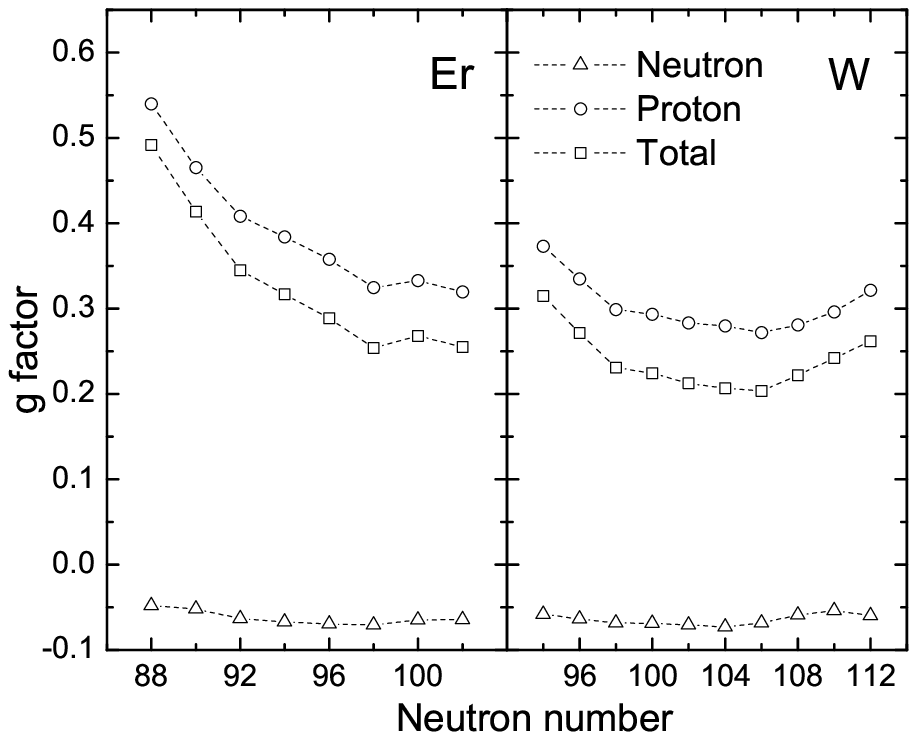}{0pt}{0pt} { Calculated first excited
$2^+$ state $g$ factors for Er and W isotopes, with decomposition
of the total g factor (open squares) into proton (open circles)
and neutron (open triangles) contributions.}

Fig. 5 shows the calculated $g(2^+_1)$ factors for Er and W
isotopes. To understand the $N$-dependent variations, theoretical
results are decomposed further into the individual contributions
from the proton and neutron operator (see Eq. (\ref{gfactor})),
the sum of which is the total $g$ factor that was compared with
data in Fig. 4. Remarkably, we find that the $g(2^+_1)$ variations
originate mostly from the proton contribution while the neutron
contribution is very small and stays nearly constant within each
of the isotopic chain. Therefore, the change of the proton
contribution alone describes the observed variation trends of
$g(2^+_1)$ factors, namely, a rapid decrease followed by a
constancy in the Er nuclei and probably a U-shape in the W nuclei
(see Fig. 5). It is rather interesting that the proton
contribution varies as the neutron number changes along an
isotopic chain. This can only be possible with the presence of
strong neutron-proton interaction. In this regard, we notice that
Zhang {\it et al.} \cite{Zhang06} suggested that the observed
constancy of the $g(2^+_1)$ factors in the well-deformed region is
attributed to the reduction of proton-neutron interaction
strengths near the midshell.

While the calculation predicts a flat behavior of $g(2^+_1)$ for
the lighter Os nuclei, a clear departure from data occurs for
heavier Os isotopes and for all the Pt isotopes considered in this
paper. Although for $^{198}$Pt, the theoretical value becomes
closer to the data, we must conclude that the present calculation
fails to describe the observed $g(2^+_1)$ trend in the Os and Pt
chains. We have tried various calculations by constructing our
deformed model space with different $\epsilon_2$ deformations. In
the example of $^{192}$Pt, it is found that with artificially
increasing basis deformation, the $g(2^+_1)$ factor values starts
going up, and at $\epsilon_2=0.24$ the calculated $g(2^+_1)$
agrees with data. However, the experimental energy levels and
$B(E2)$ values in this nucleus cannot be simultaneously described.
This may indicate that, although g-band energies and $B(E2)$s in
the Os and Pt chains are reproduced by the model, the obtained
wavefunctions with respect to the single-particle content can be
wrong. We note that energy levels and $B(E2)$s near the ground
state reflect mainly the collective properties of even-even nuclei
and are not sensitive to single particles.

This calls for a further improvement of the projected shell model
type approaches to generally describe transitional nuclei. More
correlations in the wavefunction need to be included, which goes
beyond what an axially deformed quasiparticle vacuum state can
contain. These correlations can be introduced by the addition of
the $D$-pair operators to the vacuum state \cite{Sun03}, which
takes both quasiparticle and collective degrees of freedom
explicitly into account in a shell model basis. The generator
coordinate method, which consists of a construction of a linear
superposition of different product wave-functions, can also be
adopted.

\section{Summary}

Inspired by the recent experimental work of Berant {\it et al.,}
\cite{Berant04}, we have made an attempt to study systematically
$g(2^+_1)$ for all the isotopes from Gd to Pt, using the projected
shell model approach. With a single set of interaction strengths,
we have carried out calculations for each nucleus in a projected
basis constructed with appropriate deformations. We have been able
to reproduce the energy levels and $B(E2)$s for low-lying states
in the g-band for all the 61 nuclei considered in the paper. With
the same set of calculation conditions, we have calculated $g$
factors of the first $2^+$ state. It has been found that for the
isotopes from the Gd to W chain, the characteristics of
experimental data along each isotopic chain are described by the
theoretical calculations, such as the downsloping trend in the Gd,
Dy, and Er isotopes, the upsloping trend in the W isotopes, and
the flat behavior of the Yb and Hf isotopes. For the heavier Os
and Pt nuclei, the results have indicated that although the energy
levels and $B(E2)$s can be described equally well as in the
lighter nuclei, the calculated $g$ factors are wrong. Study of the
separate contributions of proton and neutron to the $g(2^+_1)$
factors suggests that the variations of the $g$ factors as the
neutron number changes originate mainly from the proton
contribution. The overly weak proton contribution to the $g$
factors for Os and Pt isotopes indicates deficiency in the
wavefunctions. To describe the heavier isotopes in the Os and Pt
chains, improvement in the theory is required.

\section{acknowledgments}

Y.S. thanks the colleagues at the Physics Departments of Xuzhou
Normal University and Tsinghua University for the warm hospitality
extended to him. This work is partly supported by the China NNSF
with grant 10325521, and by the NSF of USA under contract
PHY-0140324 and PHY-0216783.

\baselineskip = 14pt
\bibliographystyle{unsrt}

\end{document}